%Paper: hep-ph/9501306
%From: masiero@ipdgr4.pd.infn.it
%Date: Mon, 16 Jan 1995 13:09:21 +0200

\magnification=1200
\hsize=15truecm
\vsize=23truecm
\baselineskip 18 truept
\voffset=-0.5 truecm
\parindent=1cm
\overfullrule=0pt

\def\lqua{\lower4pt\hbox{\kern5pt\hbox{$\sim$}}\raise1pt
\hbox{\kern-8pt\hbox{$<$}}~}

\centerline
{\bf MIXED DARK MATTER AND THE FATE OF BARYON}
\smallskip
\centerline
{\bf AND LEPTON SYMMETRIES $^*$}

\vskip 1truecm

\centerline
{\sl A. Masiero}

\centerline
{\sl Istituto Nazionale di Fisica Nucleare}
\centerline
{\sl 35131 Padova (Italy)}

\smallskip

\centerline
{\sl and}

\smallskip

\vskip 0.5truecm

\centerline
{\sl Dipartimento di Fisica - Univ. di Perugia}
\centerline
{\sl 06100 Perugia (Italy)}

\vskip 1truecm

$^*$ Invited talk given at the XVIII Johns Hopkins Workshop, September
1994, Florence, Italy.

\vfill\eject
\noindent
{\bf Abstract.} The available data on large scale structures seem
to favour models with mixed dark matter (MDM), i.e. with a hot and
cold component  in a rather well--defined amount, or with some form of
``warm" dark matter. I discuss some prospects for these new scenarios
for DM in the context of supersymmetric extensions of the electroweak
standard model. In particular, I emphasize the intriguing link which
exists between the present prospects of solution of the DM puzzle and
the explicit or spontaneous breaking of baryon and/or lepton number
symmetries. Some consequences on the issue of baryogenesis are worked out.

\vskip 0.5truecm

\noindent{\bf 1.\ INTRODUCTION}

\vskip 0.3truecm

All the three branches which merge together into the relatively recent
field of astroparticle physics exhibit a standard model. In particle
physics this is the extraordinarily successful Glashow--Weiberg--Salam
description of electroweak interactions, in astrophysics we have the
standard picture of stellar evolution and in cosmology the hot Big Bang
model represents our standard view of the early Universe. Obviously all
these three models have become standard thanks to numerous and solid
experimental pieces of evidence. In particular, let me remind some particle
physicists who consider cosmology rather far from being experimentally
testable, that the expansion of the universe, the prediction of the
cosmic microwave background radiation
and its temperature, and the prediction of the abundances of the primordial
elements from nucleosynthesis represent solid experimental pillars which
severely constrain any attempt to propose a cosmological model. In fact
also in cosmology we are now entering a new phase of observational activity
both with large (10-m) ground telescopes and satellite activity.

Given that the three above standard models have common areas, one can
naturally wonder whether there is a full compatibility. There is a good
chance that this is not the case: the solar neutrino problem may represent
the clash between ths standard solar model and the standard GWS
model (where
neutrinos are strictly massless), while the dark matter problem may be the
hint of a severe clash between one of the most stringent predictions of
nucleosynthesis, the number of surviving baryons, and the absence in the
GWS model of relic particles which may be needed to account for the
presence of dark matter.

Indeed, I would say that at the moment the electroweak standard model
does not feel any serious threat from the accelerator data (the potential
discrepancies concerning $\Gamma(Z\rightarrow b\bar b)$, the SLAC data on
the left--right asymmetry and the semileptonic branching ratio of the B
meson may turn out to be real problems for the GWS model, but it is
certainly premature to draw any conclusion so far). In a sense, the solar
neutrino and dark matter problems may represent the only ``observational"
hint for the need of new physics beyond this model.

In this talk I would like first to discuss to what extent the dark matter
problem actually calls for new physics (sect. II). Then, I'll turn to
analyze which kind of new physics may be more suitable for the solution
of the DM problem and the related issue of large scale structure formation
(sect. III). Sect. IV will be devoted to a study of the relation between DM
and the most attractive extension of the SM, i.e. the minimal
supersymmetric standard model (MSSM). In Sect. V I'll deal with the
lightest supersymmetric particle as a favourite candidate for cold DM
In Sect. VI I'll introduce a new
subject which I think is intimately related to the issue of DM, namely the
violation of baryon $(B)$ and lepton $(L)$ numbers at finite temperature and
its implications for models with explicit or spontaneous breaking of $L$.
Here the link between two major issues of modern cosmology, i.e.
baryogenesis and DM, should appear in all its evidence. In particular this
study may be relevant in constraining the different options which are
present in the supersymmetrization of the SM. The most recent options for
the solution of the DM puzzle with the presence of mixed DM or warm DM will
be briefly discussed in Sect. VII.

\vskip 0.5truecm

\noindent{\bf 2.\ DOES THE DM PROBLEM CALL FOR NEW PHYSICS ? }

\vskip 0.5truecm

I think that the most relevant question for a particle physicist when
tackling the problem of DM is whether the solution of this puzzle calls for
extensions of the electroweak SM. Let us briefly state the ``facts"
$^{[1]}$.

The contribution of luminous matter to the energy density of the Universe
$\Omega=\rho/\rho_{cr}$ ($\rho_{cr}=3H^2_0/8\pi G$ where $G$ is the
gravitational constant and $H_0$ the Hubble constant) is less than 1\%. The
most solid piece of evidence that we need DM comes from the rotation curves
of spiral galaxies, with a value of $\Omega_{DM}$ in the 10\% range. Given
that in the SM the only candidates to produce all this enormous amount of
non--shining matter are baryons, one can ask: can we  account for
$\Omega_{DM}$ just using non--shining baryons ? Here comes a crucial
constraint on $\Omega_B$ from the Big Bang nucleosynthesis. The ratio of
the baryon to photon number densities is one of the three key--elements
which established the moment of start of nucleosynthesis and, hence, the
abundances of the primordial elements which are produced throughout this
process. From detailed analyses one concludes that $\Omega_B$ cannot
exceed 10\%.

On the other hand there are indications for larger values of $\Omega$ when
one applies the usual dynamical methods to scale structures at distances
larger than the galactic scale. Last, but certainly not least for a
theorist, $\Omega=1$ is predicted in inflationary models and I think that
it is completely fair to say that so far we have not any other viable way to
tackle formidable cosmological problems such as causality, oldness and
flatness of the Universe, but the inflationary path.

Clearly then, if one believes that $\Omega$ should exceed 10\% there is no
way at all to accommodate this value of $\Omega$ just involving the presence
of the surviving cosmic baryon asymmetry.

In the SM the relics of the primordial Universe  are the photons (the
famous cosmic background radiation at 2.7 $^0$K), the massless neutrinos
(with a number density slightly smaller than that of the photons) and the
surviving baryons and charged leptons. Since $\Omega_\gamma$ and
$\Omega_\nu$ are certainly much smaller than 10\%, we conclude that we need
to extend the SM to schemes with additional relic particles if we are to
explain $\Omega > 0.1$.

\vskip 0.5truecm

\noindent{\bf 3.\ WHAT DM IS MADE OF}

\vskip 0.5truecm

The first broad distinction among the several candidates for DM which have
been proposed in the literature concerns the amount of interactions of a
particle with all the others in the primeval plasma. The typical scale one
has to compare these interactions with is the expansion rate of the
Universe which vastly changes with time. Hence at a certain moment
throughout the history of the Universe a particle can exhibit interactions
whose rates are larger than the expansion rate of the Universe,
while at other times the opposite situation can occur. The former
case refers to a situation in which the particle is said to be in thermal
equilibrium, whilst in the opposite case we have a particle which is
decoupled.

There are particles whose interactions are so weak that they were never in
thermal equilibrium. The most representative of these non--thermal
candidates is the axion. In this talk I'll focus my attention on thermal
candidates, i.e. particles which were in thermal equilibrium for some time
during the early story of the Universe.

The traditional distinction one makes is between hot (HDM) and cold (CDM)
dark matter. Two examples can immediately clarify this distinction.

Consider a massive neutrino of few eV's. The weak interactions keep it in
thermal equilibrium as long as the temperature of the Universe is above
1 MeV. Below 1 MeV the neutrino decouples. Hence, at the moment it
decouples this neutrino is highly relativistic. This is the ``standard"
example of an HDM candidate. Now, let us envisage a kind of opposite
situation. Consider a supersymmetric (SUSY) extension of the SM where the
lightest SUSY particle is a neutralino of, say, 50 GeV. As we'll see in
next sect., this particle decouples when the temperature of the Universe is
much below its mass (roughly $\sim m_\chi/20$, where $m_\chi$ denotes the
mass of the lightest neutralino). Hence at the moment it decouples, this
particle is highly non--relativistic. We have here an example of CDM. A
more appropriate definition of CDM and HDM is linked to the problem of
large scale structure formation which is the subject to which I turn now.

To be a good candidate for DM it is not enough to provide $\Omega=1$, or
whichever value of $\Omega$ one prefers. Very severe constraints on the
nature of DM come from the crucial issue of the formation of large scale
structures (galaxies, clusters and superclusters of galaxies, etc.). The
theory of structure formation is linked to two key--elements: i) the shape
of the primordial density fluctuations whose evolution produces the large
scale structure that we observe today and ii) the content of matter in the
Universe, i.e. the nature of the DM. The variation of these two ingredients
leads to different predictions of the power spectrum, i.e. on the
distribution of structures at different distances.

Two types of origin for the seed of density fluctuations have been
envisaged: inflation and topological defects (cosmic strings,...). In the
inflationary scenarios quantum fluctuations of the inflation field are
changed into density fluctuations giving rise to a typical scale--invariant
fluctuations spectrum. The seed density fluctuations evolve under the
action of gravity. Hence their evolution is determined by the nature of
DM.

Two scales of importance for the evolution of the seed density fluctuations
are: $\lambda_{FS}$, the free streaming scale below which fluctuations in a
nearly collisionless component are damped due to free streaming and
$\lambda_{EQ}$, the horizon length when radiation--matter equality occurs
(this scale is important since density fluctuations of non--relativistic
matter within the horizon are suppressed during the radiation dominated
era, while they begin as the matter domination era starts).

Let us see how how our prototypes for HDM and CDM, the light massive
neutrino and the lightest neutralinos, behave in the process of formation
of large scale structures.

First I consider light $(m_\nu < 1$ MeV) stable neutrinos. If they have a
mass $ > 10^{-4}$\ eV they are non--relativistic today and their energy
density is simply given by $\rho_\nu=m_\nu n_\nu$, where $m_\nu$ denotes
their mass, while $n_\nu$ is their number density. This latter quantity can
be easily related to the photon number density $n_\gamma, n_\nu = (3/22)
g_\nu n_\gamma$, where $g_\nu$ is equal to 2 or 4 according to the Majorana
or Dirac nature of neutrinos. Then one can readily compute the contribution
to $\Omega$ due to the presence of these relic neutrinos:

$$
\Omega_\nu \equiv {\rho_\nu\over\rho_c} = 0.01\
m_\nu(eV)\ h^{-2}_0 \Bigl( {g_\nu\over 2}\Bigr)
\Bigl( {T_0\over 2.7}\Bigr)^3\ ,\eqno(1)
$$

\noindent
where $h_0$ is the present value of the Hubble parameter in units of
$100\ Km\ sec^{-1}$ $parsec^{-1}$ and $T_0$ is the temperature of the
microwave cosmic background radiation in degrees Kelvin.

The lower bound of $10^9$ years on the age of the Universe requires
$\Omega h^2_0 < 1$ and, therefore:

$$
m_\nu (eV) < 200\ g^{-1}_\nu\ eV\eqno(2)
$$

\noindent
for stable neutrinos which decouple while still relativistic (i.e.
$m_\nu < 1$ MeV).

Given that experimentally $h_0$ ranges between 0.4 and 1, it is
easy to see from (1) that neutrinos in the 10 eV range can readily yield
$\Omega_\nu$ in the interesting range 0.1--1. From this point of view,
clearly massive neutrinos would be the best candidates for DM providing
large values of $\Omega$ quite easily and with a major advantage on all
other competitors: of all the proposed DM candidates, neutrinos are the
only particles that we know to exist for sure !

However, as I said, it is not enough to provide $\Omega=1$ for a relic
particle to prove to be a good DM candidate. The other test concerns the
role it plays in structures formation. The O (10\ eV) neutrinos we are
considering are relativistic until late in the evolution of the Universe.
The $\nu$ density perturbations are wiped out below the free--streaming
scale

$$
\lambda^\nu_{FS} \simeq 40\ Mpc \Bigl( {30 eV\over m_\nu}\Bigr)
\eqno(3)
$$

\noindent
corresponding to the mass scale:

$$
m^\nu_{FS} \simeq 10^{15} M_\odot
\Bigl( {30 eV\over m_\nu}\Bigl)\ ,\eqno(4)
$$

\noindent
where $M_\odot $ denotes the solar mass. Hence the first structures to form
have dimension much larger than that of galaxies and
there is the problem to form enough ``small" structures in a scenario with
only neutrinos constituting the DM. The only solution  which may be viable
is the addition to neutrinos of some seeds for the formation of small
structures. Cosmic strings are the best known candidates to play such a
role. Whether schemes with pure HDM and cosmic strings may reproduce
correctly the known power spectrum is a highly debated issue and the
improvement of the current numerical simulations will hopefully shed some
light on this intriguing question.

The difficulties which are present in any scheme with pure HDM to account
for the structure formation made scenarios with pure CDM even more favoured
for several years. The so--called standard cold dark matter model
$^{[2]}$ predicted
$\Omega=1$, with $\Omega_{CDM}\sim 90-95\%$, $\Omega_B\sim 5-10\%$ and
$\Omega_{\nu,\gamma} < 1\%$. The seed fluctuations were generated during
inflation and with a scale--invariant spectrum. In this model
$\lambda_{EQ} \simeq 30 (\Omega h^2_0)^{-1}$\ Mpc. Although some problems
were present even before the advent of the COBE data $^{[3]}$, the situation
 has
become rather difficult for the pure CDM scenario after COBE. With the
normalization fixed at the COBE data $^{[4]}$
the CDM model predicts more power at
small scales than observed $^{[5]}$.

Several remedies have been proposed modifying either the initial fluctuation
spectrum or the composition of DM. To ``disfavour" the formation of
structures at small scales one could try to increase the above value of
$\lambda_{EQ}$. Late decaying particles $^{[6]}$
or a conspicuous contribution of
the cosmological constant to $\Omega$
(with $\Omega_{CDM}\sim 0.2$) $^{[7]}$ can
yield such an increase of $\lambda_{EQ}$. The other option to solve the
problem is obvious from our previous analysis of the virtues and faults of
HDM and CDM scenarios. Since they suffer from opposite problems when
dealing with the structure formation, one might expect that a convenient
admixture of both components may reproduce the whole power spectrum
correctly. It turns out that the best fit is provided by the $\Omega_{CDM}
\sim 0.6$ and $\Omega_{HDM} \sim 0.3$ $^{[8]}$
There has been some work along the lines of these mixed dark
matter scenarios and some aspects will be discussed in sect. VII. The other
possibility that one can envisage is to have a DM candidate which is
somewhat ``colder" than the abovementioned light neutrinos so that
$\lambda_{FS}$ can decrease. Also some example of this kind of warm DM will
be provided in sect. VII.

{}From the above discussion it emerges that at least some amount of
$\Omega_{DM}$ should be accounted for by the presence of cold dark matter.
Before the impressive results of LEP a popular candidate for CDM was a
heavy neutrino with a mass in the GeV range. Indeed one can find that
$\Omega_\nu h^2_0\sim 3$(GeV/$m_\nu)^2$ and, hence, having $m_\nu\sim$ few
GeV one could easily obtain $\Omega_\nu\simeq 1$. However if these new
heavy neutrinos couple to the $Z$ boson in the same way ordinary neutrinos
do, they would contribute too much to the $Z$ invisible width.
The only way to drastically reduce this contribution is if these neutrinos
have masses close to $m_Z/2$, but in this case $\Omega_\nu$ drops down to
$0(1\%)$ making these neutrinos uninteresting for the DM problem.

The favoured CDM candidate has to do with what I consider the most
``plausible" extension of the SM, i.e. its supersymmetrization. This is
the issue that I intend to discuss in the next sect.

\vskip 0.5truecm

\noindent{\bf 4.\ DARK MATTER AND SUPERSYMMETRY}

\vskip 0.5truecm

There are several reasons which favour the presence of supersymmetry
(SUSY) among the fundamental symmetries $^{[9]}$.
In my view the most compelling
one is related to the incorporation of gravity with the other three
fundamental interactions through supergravity. However, for that matter
supersymmetry might as well be a good symmetry at the Planck scale being
broken below that scale. If that is the case, then we should not
bother so much about SUSY from the phenomenological point of view. What is
actually crucial for the TeV physics to be tested in the coming machines
is that supersymmetry has to be present much below the Planck scale, indeed
down to the electroweak scale of $0(10^2-10^3$\ GeV), if we are to
invoke  supersymmetry to alleviate the gauge hierarchy problem. As is well
known, this problem is related  to the presence of fundamental scalar
particles in the SM. The most radical cure for the problem would be the
elimination altogether of elementary scalars, but then one has to envisage
some kind of dynamical mechanism for the spontaneous breaking of the
electroweak symmetry. Since so far no consistent model of this kind has
been proposed (in spite of years of relentless efforts along these lines),
low energy supersymmetry (meaning SUSY extensions of SM with SUSY broken
only at $10^2-10^3$\ GeV) represents the only consistent way we have at the
moment to cope with the gauge hierarchy problem.

A point of utmost relevance which is often forgotten when discussing the
supersymmetrization of the SM is that there is no unique way to realize
a SUSY version of the SM. The simplest thing one can try is to use just the
fields of the SM enbedding them into the convenient superfields and then
impose the $SU(3)\times SU(2)\times U(1)\times$ SUSY symmetry. If one just
follows this kind of ``minimal prescription", the model which results is
going to be immediately ruled out for a very good reason: your protons
would have already decayed before you end reading this sentence ! Indeed,
one can construct renormalizable operators which violate either baryon (B)
or lepton (L) number in the part of the SUSY lagrangian which is known as
the superpotential. The latter constitutes a kind of SUSY version of the
ordinary Yukawa lagrangian of the SM, but with a major difference: since in
the SUSY version there exist scalar SUSY partners which carry $B$ or $L$ it
is possible to construct operator of dimension 4 containing two ordinary
fermions and one $s$--fermion which respect the $SU(3)\times SU(2)\times
U(1)$ symmetry. For instance $u_R d_R \tilde d_R$ and $u_L e_L \tilde
d^c_L$ violate $B$ and $L$, respectively $(\tilde d_R$ and $\tilde d^c_L$
denote the scalar partner of the right--handed down quark or, equivalently,
of the left--handed $Q=+1/3$ down anti--quark). Their simultaneous presence
leads to a proton decay through a 4--quark  operator mediated  by the
exchange of a down $s$--quark. Since SUSY is bound to be broken at a scale
which cannot significantly exceed 1 TeV, we would have an essentially
immediate proton decay.

The simplest possibility to avoid the above catastrophe is the addition to
the $SU(3)\times SU(2)\times U(1)\times N=1$ SUSY invariance of a new
discrete symmetry which forbids all the $B$ \underbar{and} $L$ violating
operators of the superpotential. This is the famous discrete matter
$R$--parity which assigns +1 to all kown particles of SM and -1 to their
SUSY partners. Obviously, then, no operator with two ordinary fermions and
one $s$--fermion can survive.

This situation that we encounter when supersymmetrizing the SM is
profoundly different from what occurs in the SM itself. In this model $B$
and $L$ are automatic symmetries of the theory, namely given the $SU(3)
\times SU(2)\times U(1)$ invariance and the usual field assignment it is
impossible to construct renormalizable operators which violate $B$ or $L$.

$R$--parity eliminates all operators which violate $B$ or $L$.
However, to prevent proton decay it is enough to forbid either $B$ or $L$
violation. Hence, one might wonder whether $R$--symmetry can be replaced by
other discrete symmetries which forbid either the $B$-- or the
$L$--violating renormalizable operators, but not all of them. An exhaustive
search for all these symmetries was accomplished in ref. [10].

If one imposes the stringent constraint that the $Z_n$ discrete symmetries
which accomplish the task to stop proton decay be ``discrete anomalous
free", then one is left with only two candidates: the well--known
$R$--symmetry and baryon--parity, a discrete symmetry which forbids the
$B$ violating operators, but allows for the $L$ violating ones. I'll
discuss some aspects of $B$--parity in relation to the DM problem in next
section.

There is a major implication for the DM issue if one imposes the
$R$--parity: as long as this symmetry is unbroken the lightest SUSY
particle (LSP) is absolutely stable. One can expect that together with
$\gamma,\nu$ and baryons also the LSP will be part of the relics of the
early Univers in SUSY versions of the SM with $R$--parity.

\vskip 0.5truecm

\noindent{\bf 5.\ THE LIGHTEST SUPERSYMMETRIC PARTICLE (LSP)}

\vskip 0.5truecm

In models where a discrete symmetry, matter $R$--parity $^{[9]}$ discriminates
between ordinary and SUSY particles, the lightest SUSY particle (LSP) is
absolutely stable. For several reasons the lightest neutralino is the
favourite candidate to be the LSP fulfilling the role of CDM $^{[11,12]}$.

The neutralinos are the eigenvectors of the mass matrix of the four neutral
fermions partners of the $W_3, B, H^0$ and $H^0_2$. There are four
parameters entering this matrix: $M_1, M_2, \mu$ and $tg\beta$. The first
two parameters denote the coefficient of the SUSY breaking mass terms $\bar
B\bar B$ and $\bar W_3\bar W_3$ respectively, $\mu$ is the coupling of the
$H_1-H_2$ term the superpotential. Finally $tg\beta$ denotes the ratio of
the $VEV's$ of the $H_2$ and $H_1$ scalar fields

In general $M_1$ and $M_2$ are two independent parameters, but if one
assumes that a grand unification scale takes place, then at the grand
unification $M_1= M_2 = M_3$, where $M_3$ is the gluino mass at that
scale. Then at $M_w$ one obtains:

$$
M_1 = {5\over 3} tg^2\theta_w M_2 \simeq {M_2\over 2},\quad
M_2 = {g^2_2\over g^2_3} m_{\tilde g} \simeq m_{\tilde g}/3\ ,\eqno(5)
$$

\noindent
where $g_2$ and $g_3$ are the $SU(2)$ and $SU(3)$ gauge coupling constants,
respectively.

The relation (5) between $M_1$ and $M_2$ reduces to three the number of
independent parameters which determine the lightest neutralino composition
and mass: $tg\beta, \mu$ and $M_2$. Hence, for fixed values of $tg\beta$
one can study the neutralino spectrum in the $(\mu, M_2)$ plane. The major
experimental inputs to exclude regions in this plane are the request that
the lightest chargino be heavier than $M_Z/2$ and the limits on the
invisible width of the $Z$ hence limiting the possible decays $Z\rightarrow
\chi\chi,\chi\chi'$.

Moreover if the GUT assumption is made, then the relation between $M_2$
and $m_{\tilde g}$ implies a severe bound  on $M_2$ from the experimental
lower bound on $m_{\tilde g}$ of CDF (roughly $m_{\tilde g} > 120$\ GeV,
hence implying $M_2 > 40$ GeV). The theoretical demand that the electroweak
symmetry be broken radiatively, i.e. due to the renormalization effects on
the Higgs masses when going from the superlarge scale of supergravity
breaking down to $M_W$, further constrains the available $(\mu, M_2)$
region.

The first important outcome of this is that the lightest neutralino mass
exhibits a lower bound of roughly 10 to 20 GeV $^{[13]}$. The prospects for an
improvement of this lower limit at LEP 200 crucially depends on the
composition of $\chi$ $^{[13]}$. If $\chi$ is mainly a gaugino, then it is
difficult to go beyond 40 GeV for such a lower bound, whilst with a $\chi$
mainly higgsino the lower bound can jump up to $m_\chi > M_W$ at LEP 200.

Let us focus now on the role played by $\chi$ as a source of CDM. $\chi$ is
kept in thermal equilibrium through its electroweak interactions not only
for $T > m_\chi$, but even when $T$ is below $m_\chi$. However for
$T < m_\chi$ the number of $\chi's$ rapidly decrease because of the
appearance of the typical Boltzmann suppression factor $\exp (-m_\chi/T)$.
When $T$ is roughly $m_\chi/20$ the number of $\chi$ diminuished so much
that they do not interact any longer, i.e. they decouple. Hence the
contribution to $\Omega_{CDM}$ of $\chi$ is determined by two parameters:
$m_\chi$ and the temperature at which $\chi$ decouples $(T_D)$. $T_D$
fixes the number of $\chi's$ which survive. As for the determination of
$T_D$ itself, one has to compute the $\chi$ annihilation rate and compare
it with the cosmic expansion rate $^{[11]}$.

Several annihilation channels are possible  with the exchange of different
SUSY or ordinary particles, $\tilde f, H,Z,$ etc. Obviously the relative
importance of the channels depends on the composition of $\chi$. For
instance, having assumed $\chi$ to be a pure gaugino in the case discussed
in the previous section, then the $\tilde f$ exchange represents the
dominant annihilation mode.

Quantitatively $^{[14]}$, it turns out that if $\chi$ results from a large
mixing of the gaugino $(\tilde W_3$ and $\tilde B$) and higgsino $(
\tilde H^0_1$ and $\tilde H^0_2)$ components, then the annihilation is too
efficient to allow the surviving $\chi$ to provide $\Omega$ large enough.
Typically in this case $\Omega < 10^{-2}$ and hence $\chi$ is not a good
CDM candidate. On the contrary, if $\chi$ is either almost a pure higgsino
or a pure gaugino then it can give a cospicuous contribution to $\Omega$.

As I already mentioned in the previous section, in the case $\chi$ mainly a
gaugino (say at least at the 90\% level), what is decisive to establish the
annihilation rate is the mass of $\tilde f$. LEP 200 will be able,
hopefully, to test slepton masses up to $M_W$. If there exists a $\tilde l$
with mass $< M_W$ then the $\chi$ annihilation rate is fast and the
$\Omega_\chi$ is negligible. On the other hand, if $\tilde f$ (and hence
$\tilde l$, in particular) is heavier than 150 GeV, the annihilation rate
of $\chi$ is sufficiently suppressed so that $\Omega_\chi$ can be in the
right ballpark for $\Omega_{CDM}$. In fact if all the $\tilde f's$ are
heavy, say above 500 GeV and for $m_\chi << m_{\bar f}$, then the
suppression of the annihilation rate can become even too efficient yielding
$\Omega_\chi$ unacceptably large. In conclusion if a slepton is found
at LEP 200, then the $\chi$ pure gaugino is excluded as a candidate for
CDM. If $m_{\bar f}$ is in the range 150 GeV to 500 GeV for $\chi$ in the
20 to 100 GeV range it is possible to give rise to an acceptable value of
$\Omega_{CDM}$.

Let us briefly discuss the case of $\chi$ being mainly a higgsino. If the
lightest neutralino is to be predominantly a combination of $\tilde H^0_1$
and $\tilde H^0_2$ it means that $M_1$ and $M_2$ have to be much larger
than $\mu$. Invoking the relation (5) one concludes that in this case we
expect heavy gluinos, typically in the TeV range. As for the number of
surviving $\chi's$ in this case, what is crucial is whether $m_\chi$ is
larger or smaller than $M_W$. Indeed, for $m_\chi > M_W >$ the annihilation
channels $\chi\chi \rightarrow WW, ZZ, t\bar t$ reduce $\Omega_\chi$ too
much. If $m_\chi < M_W$ then acceptable contributions of $\chi$ to
$\Omega_{CDM}$ are obtainable in rather wide areas of the $(\mu - M_z)$
parameter space. Once again I emphasize that the case $\chi$ being a pure
higgsino is of particular relevance for LEP 200 given that in this case
$\chi$ masses up to $M_W$ can be explored.

In the minimal SUSY standard model there are five new parameters in
addition to those already present in the non--SUSY case. Imposing the
electroweak radiative breaking further reduces this number to four.
Finally, in simple supergravity realizations the soft parameters $A$ and
$B$ are related. Hence we end up with only three new, independent
parameters. One can use the constraint that the relic $\chi$ abundance
provides a correct $\Omega_{CDM}$ to restrict the allowed area in this
3--dimensional space. Or, at least, one can eliminate points of this space
which would lead to $\Omega_\chi > 1$, hence overclosing the Universe. For
$\chi$ masses up to 150 GeV it is possible to find sizable regions in the
SUSY parameter space where $\Omega_\chi$ acquires intersting values for the
DM problem. A detailed and updated analysis is presented in ref. [15]
where one can compare the allowed SUSY parameters area with or without the
constraint $0.1 < \Omega_\chi h^2 < 0.7$, where $h$ is the Hubble
parameter.

There is a further phenomenological constraint which helps in restricting
even more severely the available regions of SUSY parameter space where
$\Omega_\chi h^2$ can be relevant for the DM problem: it is the recent
measurement of the decay $b\rightarrow s+\gamma$ at the inclusive level by
the CLEO collaboration. Two papers $^{[16]}$  have recently thoroughly
 investigated the
problem of the direct detection of relic neutralinos in processes of
neutralino--nucleus scattering including the constraint arising from the
experimental result of $BR(b\rightarrow s+\gamma)$. It turns out that large
portions of the SUSY parameter space where it would be possible to have a
neutralino--nucleus scattering rate high enough to be detectable in the
next round of experiments predict very large values for $BR(b\rightarrow
s+\gamma)$ vastly exceeding the experimental result. However, there still
survive particular regions where rates as high as $10^{-1}$ events/kg/day
for a $^{76}Ge$ detector are allowed. This is the case, for instance, for
relatively large $\tan\beta$ $(\tan\beta \sim 20)$ and moderate values of
the SUSY parameters $(\tilde m=200$ GeV, $\mu=-300$\ GeV, $M_Z=100$\ GeV).
For a complete discussions I refer the interested reader to the works of
ref. [16].

I close this section with a remark concerning the possibility that gauginos
are massless, i.e. $M_1=M_2=M_3=0$, to start with and that $R$--invariance
(the continuous $U(1)$ symmetry associated with the fermionic partners of
the gauge bosons, not to be confused with the discrete $R$--parity) is
broken spontaneosuly by Higgs $VEV's$ or else explicitly by dimension 2 or 3
SUSY--breaking terms in the low energy effective lagrangian. Gluino and
lightest neutralino masses then depend on only a few parameters. For a
breaking scale of a few hundred GeV or less, the gluino and the lightest
neutralino have masses typically in the range $10^{-1}-2$ GeV. On the other
hand, for a SUSY--breaking scale several TeV or larger, radiative
contributions can yield gluino and lightest neutralino masses of O(50--300)
GeV and O(10--30) GeV, respectively. As long as the Higgs $VEV's$ are the
only source of $R$--invariance breaking, or if SUSY breaking only appears
in dimension 2 terms in the effective lagrangian, the gluino is generically
the lightest SUSY particle, hence modifying the usual phenomenology (and in
particular the conventional view of the DM in SUSY) in interesting ways. For
reasons of space I cannot deal more with this interesting (or at least
curious) issue here and I recommend in particular sect. 5 of our
paper $^{[17]}$ with
G. Farrar for hints at how the DM problem may be affected by the initial
presence of a continuous $U(1)$ $R$--symmetry in supergravity models.

\vskip 0.5truecm

\noindent{\bf 6.\ LEPTON NUMBER VIOLATION IN SUSY}

\vskip 0.5truecm

In the previous section I discussed the more conventional SUSY schemes
where $R$ parity is imposed to avoid all the $B$ and $L$ violating
operators in the superpotential. From the cosmological point of view the
most important consequence of the presence of $R$ is that there exists a
stable SUSY particle which has good chances to constitute the CDM in an MDM
scenario. As for the hot part of the MDM one can think of neutrinos getting
a small mass (in the eV range). In some SUSY GUT's like SO(10) this is
naturally achieved through a see--saw mechanism.

Let me comment now the alternative possibility that $R$--parity is replaced
by some other symmetry, for instance $B$--parity, allowing for $B$ or $L$
explicit violation in the superpotential. The removal of $R$--parity has an
unpleasant consequence for  the DM problem: we lose our beloved CDM
candidate represented by the stable LSP. In models with broken $R$--parity
the LSP can decay into ordinary particles and, generally, these decays are
much faster than what would be required to make the LSP survive until
today.

The only exceptions are situations of extremely tiny violations of
$R$--parity. An example is offered in ref. [18]. Not only can the lightest
neutralino still be the CDM today, but its slow decays can have an
experimental impact: for instance, we considered the possibility of the LSP
radiative decays into a $\nu+\gamma$ with a possibly ``visible" neutrino
line. The negative result of a search performed at Kamiokande of such
neutrinos led to a sharp improvement $^{[19]}$ on the bounds of the LSP
lifetime (it
turns out that $\tau_{LSP}$ must exceed the Universe lifetime by several
orders of magnitude).

Although the absence of $R$ parity carries the bad news that in general we
lose the obvious SUSY candidate for CDM, it can have a positive impact on
the other side of a mixed dark matter (MDM) scenario, i.e. it can yield a
good amount of HDM. The point is that $R$ violation is accompanied by $L$
violation (for instance in schemes with $B$--parity), hence allowing for
nonvanishing  neutrino (Majorana) masses. In addition to the presence of
$m_\nu$ there are several other important astrophysical implications:
possibly large neutrino magnetic moments, new features in the
implementation of the MSW mechanism for the solar neutrino problem, etc.
$^{[20]}$.

The explicit violation of $L$ through the presence of $L$ violating
operators in the superpotential is severely limited not so much by
phenomenological constraints $^{[21]}$ , but rather by a powerful cosmological
argument related to the survival of the cosmic matter--antimatter
asymmetry $^{[22]}$.

The argument goes as follows. It is well--known that  owing to the
anomalous character of the $L$ and $B$ currents, these two numbers are
violated at the quantum level. Only the combination $B-L$ is conserved.
Although these violations are unlikely to produce any visible effect at
zero temperature, they become quite relevant at high temperature
$^{[23]}$: the
associated $B$ and $L$ violating processes have rates larger than the
expansion rate of the Universe (at least for 100 GeV $< T <$ critical
temperature of the electroweak phase transition, but, presumably, also for
$T > T_c$), hence leading to an equal erasement of the pre--existing $B$
and $L$ asymmetries. Hence, if one starts with $\Delta B = \Delta L$, which is
the case in GUT's with $B-L$ conservation, one ends up with $\Delta B=
\Delta L=0$ at the electroweak phase transition.

Whether these same quantum effects which are responsible for the cosmic
$\Delta B$ erasement can be used to produce a new $\Delta B$ at the time
of the electroweak phase transition is very doubtful. The survival of a
lately produced $\Delta B$ seems to require an excessively light Higgs
boson in the SM and also the amount of CP violation is unlikely to be
sufficient to obtain a sizeable $\Delta B$. However, both these objections
are far from being settled and further work is needed to make some final
assessment on this intriguing issue. A safer way to solve problem is
represented by a different boundary condition at the GUT scale with $\Delta
B\not =\Delta L$. If this is the case, given that quantum effects preserve
$B-L$ it is never possible to reach a total erasement of $\Delta B$. This
is the reason why models like SO(10) where $B-L$ is violated (hence
allowing for $\Delta B\not =\Delta L$) are certainly favoured with respect
to GUT's with $B-L$ conservation (like SU(5)).  Moreover SO(10) schemes can
lead to neutrino masses in the convenient range to provide viable
candidates for HDM.

All what I said above holds provided that during the interval time from the
production of the cosmic $\Delta B$ (for example at the GUT time) down to
the electroweak phase transition no other $B$ or $L$ violating interaction
is in equilibrium apart from the abovementioned anomalous quantum effects.
For instance, if $R$ violating processes are present and are fast enough to
be in equilibrium at some moment, since they violate either $B$ or $L$ they
certainly violate $B-L$ and hence no combination of $\Delta B$ and $\Delta
L$ can survive (independently from whether $\Delta B=\Delta L$ or $\Delta
B\not = \Delta L$ to start with). Requiring the $R$--violating induced
processes to be out of equilibrium places such a severe bound $^{[22]}$
on the
strength of the $R$ violation in the superpotential that certainly one
could forget about any phenomenological implication of $R$ breaking. As
usual, however, this is not the end of the story concerning SUSY models
without $R$ parity. Several solutions have been pointed out to let $\Delta
B$ survive even in the presence of non--negligible $R$--breaking effects.
Nervertheless the above cosmological observation represents a severe
warning for the construction of consistent SUSY schemes which are
alternative to those with the traditional matter $R$--parity.

One final comment on $R$--parity breaking is in order. We know that many
continuous or global symmetries of the initial lagrangian can be
spontaneously broken. One might wonder whether $R$--parity can undergo a
similar destiny. Long ago it was pointed out $^{[25]}$
that there are regions of the
SUSY parameter space where the minimization of the scalar potential leads
to a nonvanishing $VEV$ for the scalar partner of the neutrino, the
sneutrino. This would correspond to the spontaneous breaking of $L$ and
$R$--parity. By now we know that such a breaking is phenomenologically
forbidden. Indeed, the $Z$  boson could decay into the Goldstone boson
associated to the breaking of $L$ and the scalar partner of it. The
stringent bound on the invisible width of the $Z$ excludes
this possibility.\footnote*{It was recently discussed the possibility that
gravitational effects spoil any global symmetry $^{[26]}$.
If this is the case, $L$
might be explicitly broken very tinily. The subsequent ``spontaneous" breaking
through a $VEV$ of the sneutrino gives rise to a pseudo--Goldstone boson.
Interestingly enough, even though the explicit breaking is very small, the
mass of this particle can easily exceed the $Z$ mass hence preventing the
abovementioned decay which contributed to the $Z$ invisible width $^{[27]}$.}

Alternatively one can supplement the usual particle spectrum of the minimal
SUSY model with one or more gauge singlet scalar superfields which carry
$L$ and acquire a $VEV$ $^{[28,29]}$.
In this case the Goldstone boson being a gauge
singlet does not couple to the $Z$ boson. In
relation to the above considerations on baryogenesis and $R$--breaking, it
is relevant to notice that the breaking of $R$ can be induced radiatively,
i.e. by the evolution of the singlet masses dictated by the renormalization
group equations. It was recently shown  $^{[29]}$ that this radiative breaking
 can
delay the breaking of $R$ down to temperature so low that the $B$ violating
quantum effects are no longer effective, i.e. typically $T< 100$ GeV.

\vskip 0.5truecm

\noindent{\bf 7.\ MIXED AND WARM DM}

\vskip 0.5truecm

As discussed in the Introduction, schemes with pure hot DM or pure cold
DM seem disfavoured by recent (and also less recent) observations. Among
the new options which are presently envisaged I think that the following
two are of particular interest for particle physicists: mixed DM (MDM) and
warm DM.

MDM $^{[30]}$ relies on a scenario where $\Omega_{CDM} \simeq 2\Omega_{HDM}
\simeq
0.6$ and $\Omega_B \lqua 0.1$. In principle one does not have to sweat so
much to realize a scheme of this kind. Take a SUSY model with $R$---parity
where neutrinos are massive. Then the lightest neutralino can play the role
of CDM, while a neutrino of few $eVs$ yields the HDM. Choosing the
parameters conveniently one can obtain the prescribed cocktail of C- and
H- DM. The problem that I see is just in this convenient choice of
parameters. This is another way to say that one actually performs a
fine--tuning to obtain the correct amount of $\Omega_{CDM}$ and
$\Omega_{HDM}$
and this is certainly unsatisfactory. This is the reason which prompted
some authors to investigate some possible common origin for HDM and CDM in
order to justify close relation of their contributions to $\Omega$. In the
work of ref. [31] it was proposed to have the relative abundances of the
HDM and CDM components set by the same scale. In their model, this is the
scale of B-L spontaneous breaking of O(1 TeV). The HDM is given by the tau
neutrino, while CDM is provided by the fermionic partner of the Goldstone
boson associated to  the B-L breaking.

Together with Bonometto and Gabbiani, we proposed $^{[32]}$
an example where one same
particle may play the twofold role of HDM and CDM.
In SUSY the axion possesses a fermionic
partner, the axino $(\tilde a)$. In fact, the $\tilde a$ is likely to be
the lightest SUSY particle. Now, axinos can be produced via two entirely
different different mechanisms in these models. First there are the axinos
which are produced with the axions and were formerly in thermal equilibrium
with the other components of the Universe, subsequently decoupling at a
temperature $< V_{PQ}$  (the Peccei--Quinn scale)
much higher than their mass. This $\tilde a$
component will be an effective CDM as only fluctuations involving masses
$\lqua 0.1 M_\odot$ will be erased at its derelativization. It was shown
that they can account for $\Omega$ close to one $^{[33]}$. This kind of
``primordial" axinos are not the only axinos surviving today. Indeed if the
$\tilde a$ is the lightest SUSY particle, all the SUSY particle must
eventually decay into it.

Calling $\chi$ the lightest neutralino, we can expect the
typical decay $\chi\rightarrow\tilde a + \gamma$ to occur through a
supersymmetrization of the ordinary $a-\gamma\gamma$ coupling.

These ``second hand" axinos can easily behave as hot dark matter,
derelativizing at a redshift $z\sim 10^4$. Accordingly, fluctuations in
such component will be erased up to a mass $\sim 10^{15} M_\odot$.

The detailed study of the conditions which make this scheme a viable MDM
scenario is presented in ref. [32]. The major ingredients are a Peccei--Quinn
scale of $O(10^{10}$ GeV), heavy sfermions in the TeV range and the
lightest neutralino being a pure gaugino.

An interesting alternative to MDM is the presence of just one DM particle
which is neither cold nor hot. This warm candidate may be represented for
instance by a sterile neutrino which is somewhat heavier but less abundant
than the usual HDM neutrinos. Clearly one must be very careful about the
contribution of these extra degrees of freedom at the time of
nucleosynthesis (they must contribute less than the equivalent of half a
neutrino species). The essential point of warm DM is that it can reduce the
damping scale corresponding to the free--streaming distance that was
previously introduced. If for an ordinary HDM neutrino this damping scale
is of $O(10^{15} M_\odot)$, for the kind of warm sterile neutrinos
discussed in ref. [34] this is lowered to $10^{13} M_\odot$ hence
increasing the power on smaller scales (typically scales 1-5 Mpc).

Another example of warm DM candidate results from the ``spontaneous"
breaking of a quasi--exact $L$ symmetry (as explained in the previous
footnote). A pseudo--Goldstone boson with a mass in the keV range and with
tiny interaction with ordinary matter has been shown $^{[35]}$ to be a suitable
candidate for warm DM.

All these  attempts of a mixed and warm DM to realize a better fit to data
at different scales are certainly interesting. However I must confess that
my overall impression is that we are far from having an appealing scenario
with some compelling reason from the particle physics point of view. In
this respect scenarios with pure CDM or pure HDM were much more
attractive. The ``canonical" final sentence that more work is needed
definitely applies very well to the present situation in this field.

\vskip 0.5truecm

\noindent
{\bf Acknowledgements}. I wish to thank my ``astroparticle" collaborators,
V. Berezinsky, S. Bonometto, D. Comelli, F. Gabbiani, G. Giudice, M.
Pietroni, A. Riotto and J.W.F. Valle.
It is also a great pleasure to thank the organizers for the nice and
fruitful atmosphere in which the meeting took place.

\vskip 0.5truecm

\noindent{\bf REFERENCES}

\vskip 0.5truecm

\noindent
[1] For a clear introduction to the ``observational" aspects of the DM
problem, see, for instance: Kolb and S. Turner, in The Early Universe,
Addison--Wesley, New York, 1990;

\noindent
Dark Matter, M. Srednicki ed. (North Holland, Amsterdam, 1989);

\noindent
J.R. Primack, D. Seckel and B. Sadolet, Ann. Rev. Nucl. Part. Sci.,
\underbar{38} (1988) 751.

\smallskip

\noindent
[2] For a recent review of CDM, see M. Davis, G. Efstathiou, C.S. Frenk and
S.D.M. White, nature \underbar{356} (1992) 489.

\smallskip

\noindent
[3] G. Efstathiou, W.J. Sutherland and S.J. Maddox, Nature \underbar{348}
(1990) 705.

\smallskip

\noindent
[4] G.F. Smoot et al., Astrophys, J. Lett. \underbar{396} (1992) L1.

\smallskip

\noindent
[5] M.S. Vogeley, C. Park, M.J. Geller and J.P. Huchra, Astroph. J.
\underbar{391} (1992) L5;

\noindent
K.B. Fischer, M. Davis, A. Strauss, A. Yahil and J.P. Huchra, Astroph. J.
\underbar{402} (1993) 42.

\smallskip

\noindent
[6] H.B. Kim and J.E. Kim, preprint SNUTP 94--48, hep--ph/9405385 (1994).

\smallskip

\noindent
[7] G. Efstathiou, W.J. Sutherland and S.J. Maddox, Nature \underbar{348}
(1990) 705.

\smallskip

\noindent
[8] E.L. Wright et al., Astrophys. J. \underbar{396} (1992) L13.

\smallskip

\noindent
[9] For a phenomenologically oriented introduction to SUSY, see: H.P.
Nilles, Phys. Rep. \underbar{110C} (1984) 1;

\noindent
H. Haber and G. Kane, Phys. Rep. \underbar{117C} (1985) 1;

\noindent
For construction of the N=1 supergravity lagrangian, see E. Cremmer, S.
Ferrara, L. Girardello and A. Van Proeyen, Phys. Lett. \underbar{B116}
(1982) 231; Nucl. Phys. \underbar{B212} (1983) 413.

\smallskip
\noindent
[10]  L. Ibanez and G. Ross, Nucl. Phys. \underbar{B368} (1992) 3.

\smallskip
\noindent
[11] J. Ellis, J.S. Hagelin, D.V. Nanopoulos, K. Olive and M. Srednicki,
Nucl. Phys. \underbar{B238} (1984) 453.

\smallskip
\noindent
[12] For reviews, see, for instance: J. Ellis, Phil. Trans. R. Soc. Lond.
\underbar{A336} (1991) 247; L. Roszkowski, proceedings of the Joint
International Lepton--Photon Symposium and Europhysics Conference on High
Energy Physics (1991), Geneve, Switzerlans; K.A. Olive, Proc. of Ten Years
on SUSY Confronting experiment, eds, J. Ellis, D.V. Nanopoulos and A.
Savoy--Navarro CERN Sept. 1992.

\smallskip
\noindent
[13] L. Roszkowski, Phys, Lett. \underbar{B252} (1990) 471; L. Roszkowski,
Proceedings of the UCLA International Conference on Trends in Astroparticle
Physics, 1990.

\smallskip
\noindent
[14] A. Bottino, V. de Alfaro, N. Fornengo, G. Mignola and M. Pignone --
preprint DFTT 37/93 (1993); M. Drees, G. Jungman, M. Kamionkowski and M.
Nojiri, Phys. Rev. \underbar{D49} (1994) 636.

\smallskip
\noindent
[15] P. Gondolo, M. Olechowski and S. Pokorski;
Proc. of the XXVI International Conference on High
Energy Physics, Dallas, August 1992.

\smallskip

\noindent
[16] P. Nath and R. Arnowitt, preprint CERN--TH--7463/94 (1994);
\noindent
F. Borzumati, M. Drees and M. Nojiri, preprint DESY  94--096 (1994).

\smallskip
\noindent
[17] G. Farrar and A. Masiero, preprint RU--94--38, hep--ph
9410401 (1994).

\smallskip
\noindent
[18] V. Berezinski, A. Masiero and J.F.W. Valle, Phys. Lett. \underbar{B266}
(1991) 382.

\smallskip
\noindent
[19] M. Mori et al., Phys. Lett. \underbar{B287} (1992) 217.

\smallskip
\noindent
[20] For a review, see A. Masiero, Nucl. Phys. B (Proc. Suppl.)
\underbar{28A} (1992) 130.

\smallskip
\noindent
[21] V. Barger, G.F. Giudice and T.Y. Han, Phys. Rev. \underbar{D40}
(1989) 2987.

\smallskip
\noindent
[22] B. Campbell, S. Davidson, J. Ellis and K.A. Olive, Phys. Lett.
\underbar{B256} (1991) 457; Astropart. Phys. 1 (1992) 77;
\noindent
W. Fischer, G.F. Giudice, R.G. Leigh and S. Paban, Phys. Lett. B258 (1991)
45,

\smallskip
\noindent
[23] V.A. Kuzmin, V.A. Rubakov and M.E. Shaposhnikov, Phys. Lett.
\underbar{B155}
(1985) 36;
\noindent
V.A. Matveev, V.A. Rubakov, A.N. Tavkhelidze and M.E. Shaposhnikov, Usp.
Fiz. Nauk \underbar{156} (1988) 253.

\smallskip
\noindent
[24] H. Dreiner and G.G. Ross, preprint OUTP--92--08; A.G. Cohen and A.E.
Nelson, Phys. Lett. \underbar{B297} (1992) 111;
\noindent
A. Masiero and A. Riotto, Phys. Lett. \underbar{B289} (1992) 73; D.
Comelli, M. Pietroni and A. Riotto, preprint SISSA 93/50/A.

\smallskip
\noindent
[25] C. Aulakh and R.N. Mohapatra, Phys. Lett. \underbar{B119} (1983) 136;

\noindent
J. Ellis, G. Gelmini, C. Jarlshog, G.G. Ross and J.F.W. Valle, Phys. Lett.
\underbar{1985} 142;

\noindent
G.G. Ross and J.W.F. Valle, Phys. Lett. \underbar{B151} (1985) 375;

\noindent
R. Barbieri, D.E. Brahm, L.J. Hall and S.D.H. Hsu, Phys. Lett.
\underbar{B238} (1990) 86;

\noindent
B. Gato, J. Le\`on, J. Per\`ez--Mercader and M. Quiros, Nucl. Phys.
\underbar{B260} (1985) 203;

\noindent
M.S. Carena and C.E.M. Wagner, Phys. Lett. \underbar{B186} (1987) 361.

\smallskip
\noindent
[26] S. Giddings and A. Strominger, Nucl. Phys. \underbar{B307} (1988) 854;

\noindent
S. Coleman, Nucl. Phys. \underbar{B310} (1988) 643;

\noindent
R. Holman, S.D.H. Hsu, T. Kephart, E. Kolb, R. Watkins and L. Widrow, Phys.
Lett. \underbar{B282} (1992) 132;

\noindent
M. Kamionkowski and J. March--Russel, Phys. Lett. \underbar{B282} (1992)
137;

\noindent
S. Barr and D. Seckel, Phys. Rev. \underbar{D46} (1992) 539;

\noindent
E. Kh. Akhamedov, Z.G. Berezhiani, R.N. Mohapatra and G. Senjanovic, Phys.
Lett. \underbar{B299} (1993) 90.

\smallskip
\noindent
[27] D. Comelli, A. Masiero, M. Pietroni and A. Riotto, Phys. Lett.
\underbar{B324} (1994) 397.

\smallskip
\noindent
[28] A. Masiero and J.F.W. Valle, Phys. Lett. \underbar{B251} (1990) 273;

\noindent
J.C. Romao, C.A. Santos and J.F.W. Valle, preprint FTUV/91--06.

\smallskip
\noindent
[29] G.F. Giudice, A. Masiero, M. Pietroni and A. Riotto, Nucl. Phys.
\underbar{B396} (1993) 243;

\noindent
M. Chaichian and R. Golzalez Felice, preprint BUTP--92/36 (1992);

\noindent
J. Umemura and K. Yamamoto, Phys., Phys. Lett. \underbar{B313} (1993) 89.

\noindent
[30] Q. Shafi and F.W. Stecker, Phys. Lett. \underbar{53} (1984) 1292;

\noindent
S.A. Bonometto and R. Valdarnini, Astroph. J. \underbar{299} (1985) L71;

\noindent
S. Achilli, F. Occhionero and R. Scaranella, Astroph. J. \underbar{299}
(1985) L77;

\noindent
J.A. Holtzman, Astroph. J. Suppl. \underbar{71} (1981) 1;

\noindent
A.N. Taylor and M. Rowan--Robinson, Nature \underbar{59} (1992) 396;

\noindent
J.A. Holtzman and J. Primack, Astroph. J. \underbar{396} (1992) 113;

\noindent
D. Yu. Pogosyan and A.A. Starobinski, CUA preprint (1993);

\noindent
A. Klypin, J. Holtzman, J. Primack and E. Rego\"os, SCIPP 92/52.

\smallskip
\noindent
[31] R.N. Mohapatra and A. Riotto, Phys. Rev. Lett. \underbar{73} (1994)
1324.

\smallskip
\noindent
[32] S.A. Bonometto, F. Gabbiani and A. Masiero, Phys. Rev. \underbar{D49}
(1994) 3918.

\smallskip

\noindent
[33] K. Rajgopal, M.S. Turner and F. Wilczeh, Nucl. Phys. \underbar{B358}
(1991) 447.

\smallskip
\noindent
[34] S. Dodelson and L. Widrow, Phys. Rev. Lett. \underbar{72} (1994) 17.

\smallskip
\noindent
[35] E. Akhemedov et al., Phys. Lett. \underbar{B299} (1993) 90;

\noindent
V. Berezinsky and J.W.F. Valle, Phys. Lett. \underbar{B318} (1993) 360.

\bye